\documentclass[journal=ancac3,manuscript=article]{achemso}

\usepackage[version=3]{mhchem} % Formula subscripts using \ce{}
\usepackage{caption} %package used in order to NOT number a figure
\author{Dennis Palagin}
\email{dennis.palagin@ch.tum.de}
\author{Karsten Reuter}
\affiliation[TUM]
{Department Chemie, Technische Universit{\"a}t M{\"u}nchen, Lichtenbergstr. 4, D-85747 Garching, Germany}

\title[$M$Si$_{20}$H$_{20}$ Aggregates]
{$M$Si$_{20}$H$_{20}$ Aggregates: From Simple Building Blocks to Highly Magnetic Functionalized Materials}

\begin{document}
%%%%%%%%%%%%%%%%%%%%%%%%%%%%%%%%%%%%%%%%%%%%%%%%%%%%%%%%%%%%%%%%%%%%%
%% The manuscript does not need to include \maketitle, which is
%% executed automatically.  The document should begin with an
%% abstract, if appropriate.  If one is given and should not be, the
%% contents will be gobbled.
%%%%%%%%%%%%%%%%%%%%%%%%%%%%%%%%%%%%%%%%%%%%%%%%%%%%%%%%%%%%%%%%%%%%%

\begin{abstract}

Density-functional theory based global geometry optimization is used to scrutinize the possibility of using endohedrally-doped hydrogenated Si clusters as building blocks for constructing highly magnetic materials. In contrast to the known clathrate-type facet-sharing, the clusters exhibit a predisposition to aggregation through double Si-Si bridge bonds. For the prototypical CrSi$_{20}$H$_{20}$ cluster we show that reducing the degree of hydrogenation may be used to control the number of reactive sites to which other cages can be attached, while still preserving the structural integrity of the building block itself. This leads to a toolbox of CrSi$_{20}$H$_{20-2n}$ monomers with different number of double "docking sites", that allows building network architectures of any morphology. For (CrSi$_{20}$H$_{18}$)$_{2}$ dimer and [CrSi$_{20}$H$_{16}$](CrSi$_{20}$H$_{18}$)$_{2}$ trimer structures we illustrate that such aggregates conserve the high spin moments of the dopant atoms and are therefore most attractive candidates for cluster-assembled materials with unique magnetic properties. The study suggests that the structural completion of the individual endohedral cages within the doubly-bridge bonded structures and the high thermodynamic stability of the obtained aggregates are crucial for potential synthetic polimerization routes $via$ controlled dehydrogenation. 


\end{abstract}


%TOC figure

\begin{figure*}
\includegraphics[width=10cm]{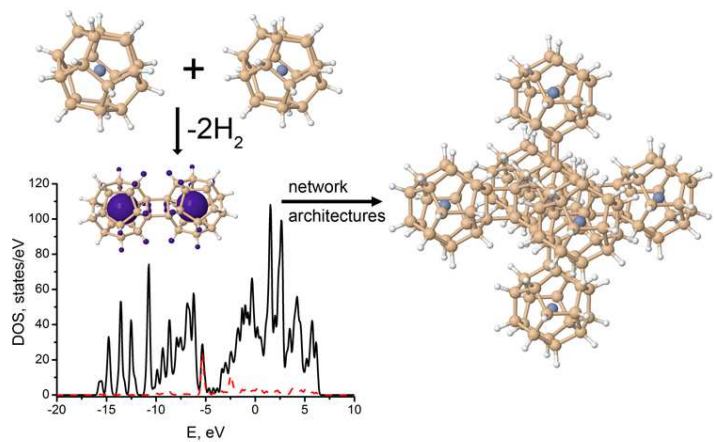}
\caption*{Table of Contents Graphic}
\label{TOC_graphic}
\end{figure*}

%\section{Introduction}

The use of multi-doped endohedral clusters as building blocks for assembly at the nanometer scale has attracted much attention due to their promising cage-like geometries and easy-to-tune electronic properties\cite{khanna09, balch12, nakajima12}. Silicon clusters are particularly suitable for such applications, since endohedral metal doping was found to produce saturated fullerene-like cage structures\cite{kumar04, khanna06, nakajima07, fielicke11, torres11}. However, the cage geometry in such structures is often stabilized through a strong cage-dopant interaction, which leads to quenching of the metal magnetic moment\cite{palagin11}. Especially interesting for magnetic applications is therefore the case of hydrogenated metal-doped Si clusters, where a minimized cage-dopant interaction preserves the high magnetic moments of the encapsulated transition metal dopants\cite{kumar03, palagin12}. Without doubt, such highly-magnetic building blocks offer great potential for instance for data storage or spin-based electronic devices\cite{schueth07, fielicke12}.

Assembling the building blocks into one material structure is, of course, a key problem in chemistry and material science. The possibility of building homo- and heterogeneous aggregations of $M$Si$_{16}$ clusters for different metal dopants $M$ has been theoretically investigated by the groups of Balb\'as\cite{torres11} and Nakajima\cite{nakajima12}. Robles and Khanna reported that assemblies of CrSi$_{12}$ clusters may have a net spin moment\cite{robles09}. However, to the best of our knowledge, no study of potentially highly magnetic aggregates of hydrogenated endohedrally doped Si cages has been performed so far. The key question in this is how such saturated clusters can be assembled at all. As we have shown in our previous work\cite{palagin12}, Si$_{20}$H$_{20}$ is the smallest Si fullerene capable of conserving a high spin state of an encapsulated 3$d$ transition metal atom, with the H termination simultaneously ensuring total saturation of all Si dangling bonds, $cf.$ \ref{fig1}a below. As opposed to this, non-hydrogenated $M$Si$_{20}$ clusters don't form cage-like structures at all as the Si compensates for its undercoordination through strong $M$-Si interaction\cite{gramzow10}. This suggests that a carefully reduced degree of hydrogenation (and, therefore, increased amount of unsaturated Si bonds) might be used to control the number of bonding sites offered by the cluster. As long as this does not jeopardize the structural integrity of the cage-like geometry, such a strategy would lead to a toolbox of monomers with differing number of "docking sites", that may offer the possibility to build network architectures of any morphology.

Scrutinizing this idea through quantitative first-principles calculations is the objective of the present study of $M$Si$_{20}$H$_{20}$ aggregates. The approach taken is as follows: First, we perform density-functional theory (DFT) based global geometry optimizations for the prototypical high-spin CrSi$_{20}$H$_{20}$ cage with different number of H vacancies introduced. This indeed validates the structural integrity of such "non-ideal" clusters. Second, we also use DFT-based global geometry optimization to show that particularly Si-Si doubly bridged dimers and trimers of these clusters actually represent most stable and highly magnetic ground-state configurations for a given Si/H composition. The energetics thus obtained finally allows us to discuss the thermodynamic feasibility of such and further polymerization to one-dimensional chains, two-dimensional sheets or, eventually, more complicated three-dimensional structures.

\section{Results and discussion}

\subsection{Building block monomers}

\begin{figure*}
\includegraphics[width=6cm]{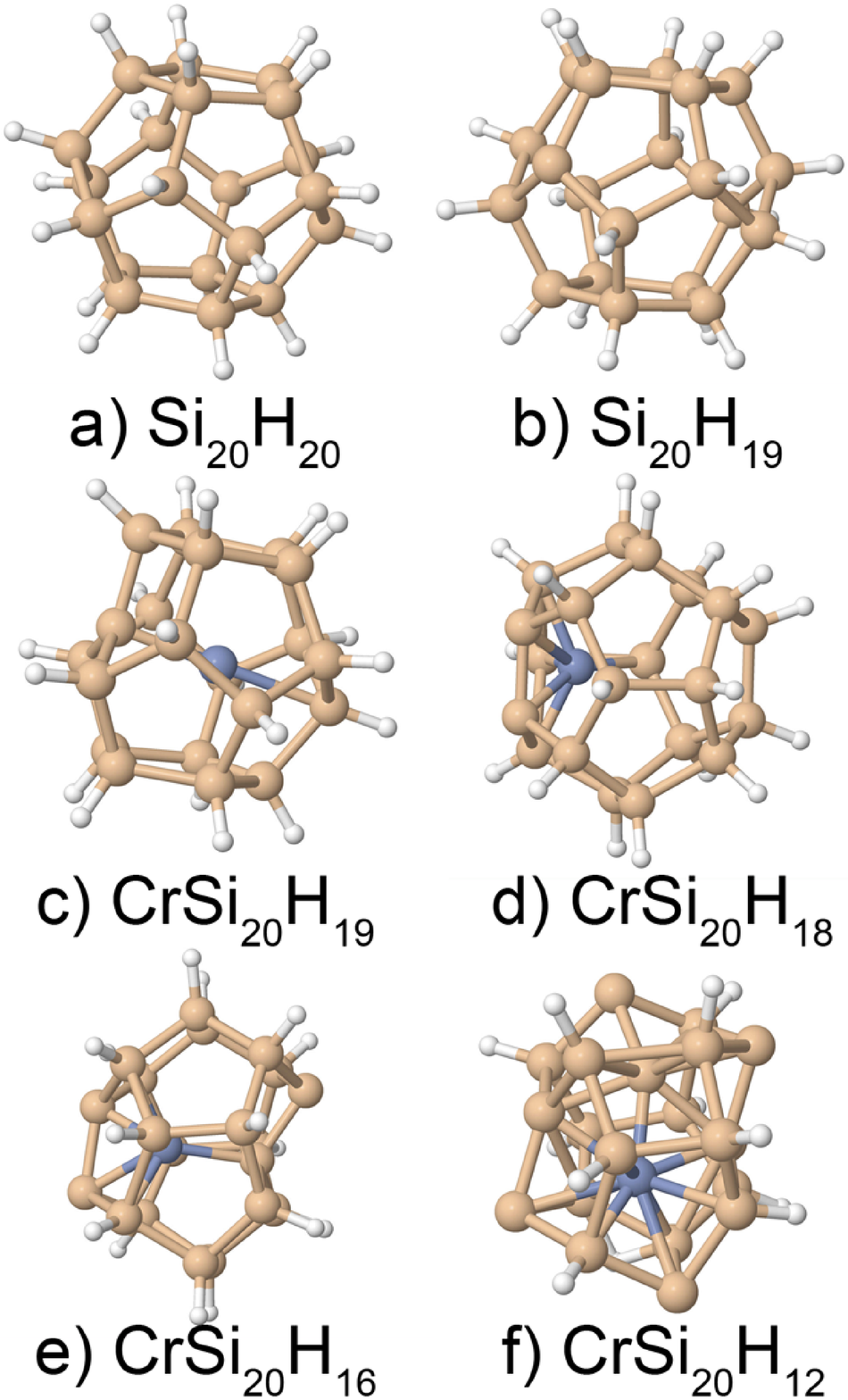}
\caption{Identified ground-state geometries of possible monomer structures, together with their spin states and energies relative to their lowest-lying non-cage isomers\cite{supplement}: a) Si$_{20}$H$_{20}$, singlet, -3.96 eV\cite{palagin12}; b) Si$_{20}$H$_{19}$, doublet, -3.18 eV; c) CrSi$_{20}$H$_{19}$, sextet, -1.23 eV; d) CrSi$_{20}$H$_{18}$, singlet, -1.95 eV; e) CrSi$_{20}$H$_{16}$, singlet, -0.80 eV; f) CrSi$_{20}$H$_{12}$, singlet, -0.24 eV.}
\label{fig1}
\end{figure*}

The starting point of our investigation is the check on the structural integrity of a $M$Si$_{20}$H$_{20}$ building block, when introducing an increasing number of H vacancies to create undercoordinated reactive Si sites that would then represent natural "docking" candidates for aggregation. At the minimized $M$-Si interaction in the hydrogenated cage, we do not expect a strong dependence on the actual metal $M$ used for doping \cite{palagin12}. This view receives support from test calculations with different metals, as well as from the fact that we obtain comparable findings also for the empty Si$_{20}$H$_{20}$ cluster, which already by itself adopts a fullerene-type cage structure, $cf.$ \ref{fig1}a. Within the focus on magnetic properties, we therefore concentrate in the following on Cr as prototypical dopant atom, which we previously reported to yield the highest septet spin state in the CrSi$_{20}$H$_{20}$ cage \cite{palagin12}.

\begin{figure*}
\includegraphics[width=10cm]{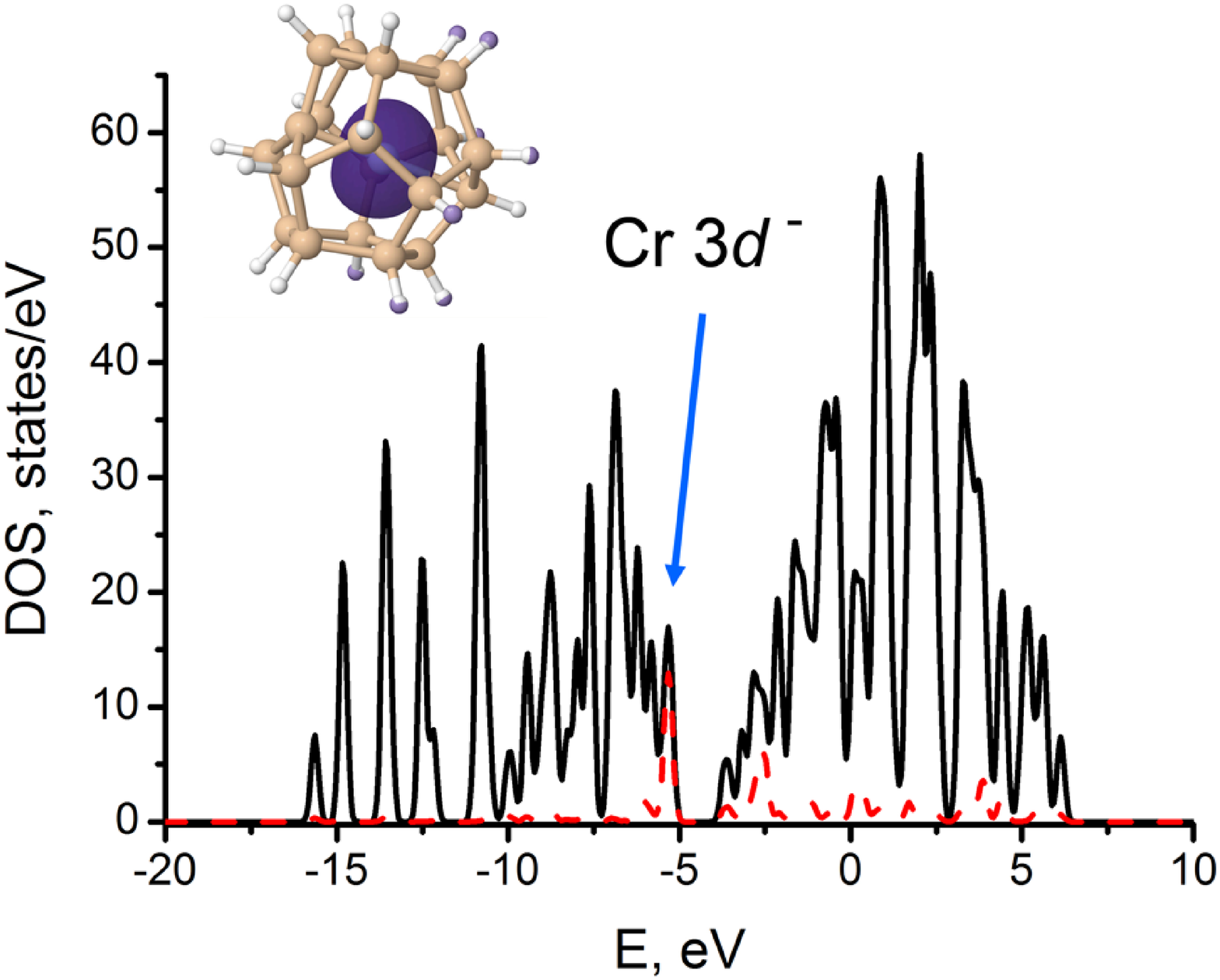}
\caption{Total density of states (DOS) (black solid line) and DOS projected on the metal dopant (red dashed line) for CrSi$_{20}$H$_{19}$ (highest occupied state lies at -5.23 eV, vacuum level is used as zero reference). The Cr-specific peak (shown with arrow) indicates that five unpaired 3$d$ electrons belong almost exclusively to the Cr atom. The inset shows the spin density distribution within the cluster, which resides almost exclusively on the metal dopant.}
\label{fig2}
\end{figure*}

Extended configurational sampling shows that one hydrogen vacancy does not change the geometrical and electronic structure much, $i.e.$ the identified CrSi$_{20}$H$_{19}$ global minimum is still a (slightly distorted) cage, $cf.$ \ref{fig1}c, with a rather high sextet spin moment. In order to further analyze the $M$-Si interaction and the concomitant magnetic properties, \ref{fig2} summarizes the total and Cr-projected density of states (DOS) of CrSi$_{20}$H$_{19}$. Consistent with the equally shown 3D spin density distribution this confirms that the unpaired electrons are predominantly located on the metal dopant, $i.e.$ the latter preserves most of its atomic character as was the case for the ideal CrSi$_{20}$H$_{20}$ cage \cite{palagin12}. 

The next lowest isomer found, which then corresponds to an irregular compact geometry, is 1.23\,eV higher in energy. For the empty Si$_{20}$H$_{19}$, $cf.$ \ref{fig1}b, this energetic gap to broken-cage geometries even amounts to 3.18\,eV. In both cases there is thus a clear thermodynamic preference for the endohedral cage, which demonstrates that the reduced hydrogenation has almost no effect on the structural integrity. What if we thus further increase the amount of H defects? \ref{fig1}d shows the identified ground-state structure for CrSi$_{20}$H$_{18}$ corresponding to two H vacancies. Again, the cage-like geometry is preserved and the energetic gap to the lowest-lying non-cage-like isomer is with 1.95\,eV quite pronounced. Still, what has changed is the position of the metal-dopant, which is now no longer more or less centered within the cage, but located close to the two dehydrogenated Si atoms. This indicates a stronger $M$-Si interaction of the type found for bare $M$Si$_{20}$ cages. This perception is confirmed by a spin DOS analysis as the one in \ref{fig2} and also reflected by the spin-quenched singlet state of the CrSi$_{20}$H$_{18}$ cage.

An intriguing feature that will become central for the aggregation discussed below is the obtained preference to locate the two H vacancies directly next to each other in the CrSi$_{20}$H$_{18}$ structure. Energetically lowest-lying alternative arrangements with the two H vacancies $e.g.$ located at opposite sides of the cage are by 0.3-0.4\,eV less favorable than the paired-vacancy ground-state isomer. This structural motif intriguingly prevails if the number of H defects is further increased. \ref{fig1}e shows the identified ground-state geometry for the CrSi$_{20}$H$_{16}$ cluster featuring a total of four hydrogen vacancies. Again, the H vacancies are paired and furthermore located at opposite ends of the cage. In fact, despite substantial sampling, this was the only H arrangement we could identify that yields a stable cage-like geometry at all. All other energetically low-lying isomers found, which start above an energetic gap of 0.8\,eV to the cage ground-state, correspond to irregular structures. Concomitant with the tendency to pair H vacancies, there is thus a tendency to arrange such H vacancy pairs as far away from each other as possible. Considering that an agglomeration of vacancies at one side of the cage is likely to break it up, the latter tendency is indirectly a consequence of the energetic stability of the cage, $i.e.$ its resilience to alternative compact geometries. Quite nicely, these effectively repulsive interactions between H vacancy pairs then allow to accommodate an unusually high amount of H defects without jeopardizing the structural integrity of the cage. This is also reflected by the obtained ground-state geometry for the CrSi$_{20}$H$_{12}$ cluster, $cf.$ \ref{fig1}f, which despite a total of eight H vacancies still maintains an (admittedly distorted, but still) endohedral geometry exhibiting four H vacancy pairs.

\subsection{Dimerization and trimerization}

\begin{figure*}
\includegraphics[width=6cm]{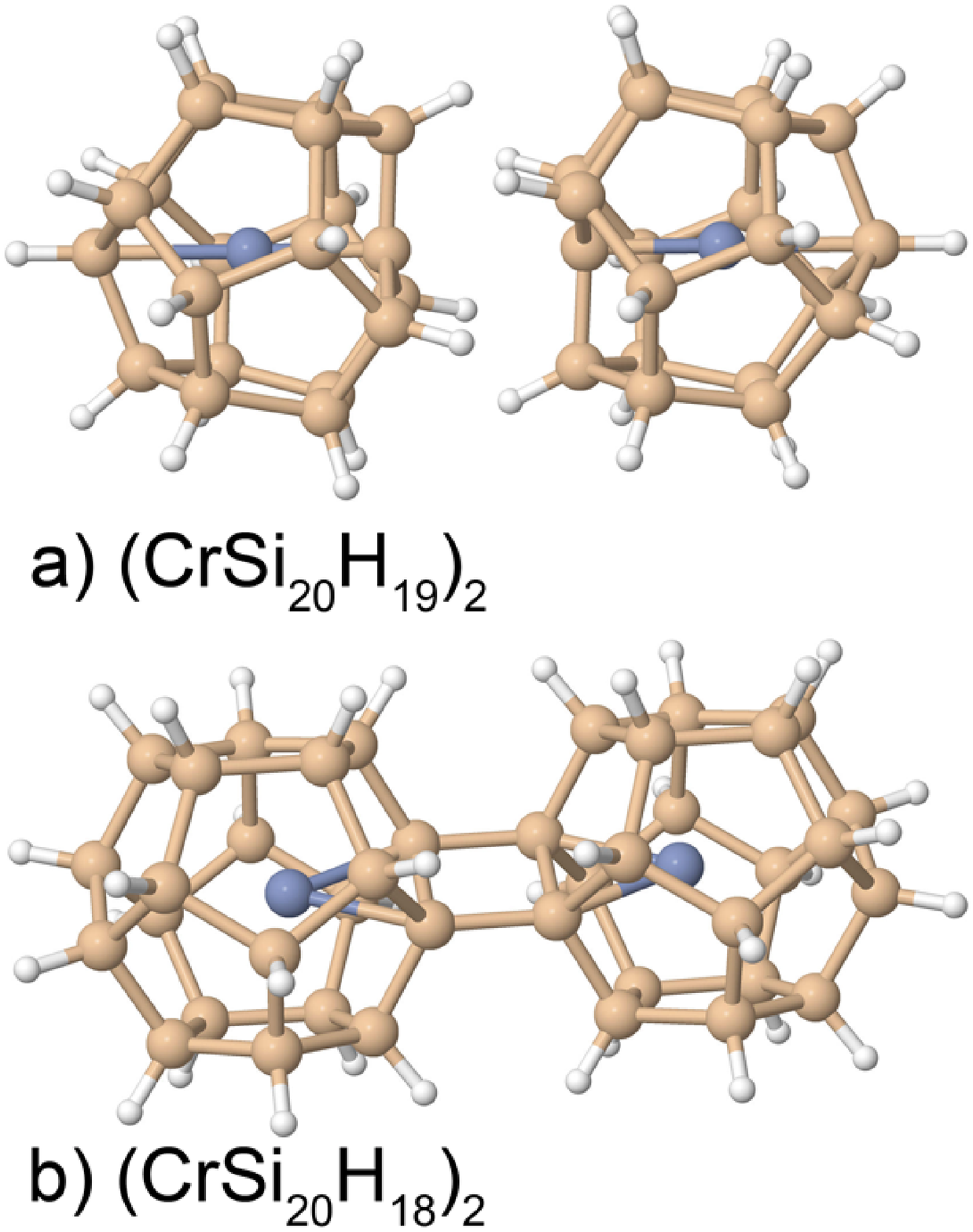}
\caption{Identified ground-state geometries of cluster dimers: a) (CrSi$_{20}$H$_{19}$)$_{2}$, ten unpaired electrons, 3.83 {\AA} intermonomer distance and 0.15 eV binding energy; b) (CrSi$_{20}$H$_{18}$)$_{2}$: twelve unpaired electrons, 2.42 {\AA} intermonomer distance and 3.46 eV binding energy.}
\label{fig3}
\end{figure*}

The results of the last section reveal the prevalence of the cage-like geometry despite an increasing degree of dehydrogenation. One would expect the H vacancy sites formed at the CrSi$_{20}$H$_{20-n}$ cages to then be natural candidates for aggregation, $i.e.$ the docking sites to fuse clusters together. With just one H defect per cluster, the obtained CrSi$_{20}$H$_{19}$ cage only allows for the formation of a dimer. However, the resulting dimer shown in \ref{fig3}a is with a computed binding energy of 0.15\,eV only very weakly bound. Notwithstanding, the geometry shown is the end result of extensive sampling runs, $i.e.$ it is the optimum structure found for a given composition of Cr$_2$Si$_{40}$H$_{38}$. In particular, this means that the identified dimer is thermodynamically stable against decomposition into any smaller sub-units (under the constraint that in the present microcanonical sampling the total number of species in all sub-units together must equal Cr$_2$Si$_{40}$H$_{38}$). Furthermore intriguing is the finding that the overall spin moment of the obtained dimer corresponds to ten unpaired electrons, $i.e.$ we arrive at a highly magnetic structure.

As reflected by the 3.83\,{\AA} distance between the undercoordinated Si atoms in the two cages, we attribute the low binding energy particularly to steric constraints that prevent the two clusters from further approaching each other without H atoms coming uncomfortably close, $cf.$ \ref{fig3}a. This should be much alleviated for the H-vacancy pairs. Indeed, we obtain a significantly increased binding energy of 3.46\,eV for the dimer formed by two CrSi$_{20}$H$_{18}$ cages, which after extensive sampling again results as ground-state isomer with as large an energetic gap as 6.63\,eV to the next low-lying structure. As apparent from \ref{fig3}b, the intermonomer distance is concomitantly significantly decreased to 2.42\,{\AA} and therewith within the normal range found for Si-Si bonds. The resulting Si double-bridge type bonding between the two monomers has recently also been suggested by Nakajima and coworkers \cite{nakajima12} as a favorable motif for the aggregation of smaller $M$Si$_{16}$ clusters. In difference to these works, the appealing feature of the present structure is that it affords for a minimized $M$-Si interaction and yields a high magnetic moment of twelve unpaired electrons.

\begin{figure*}
\includegraphics[width=16cm]{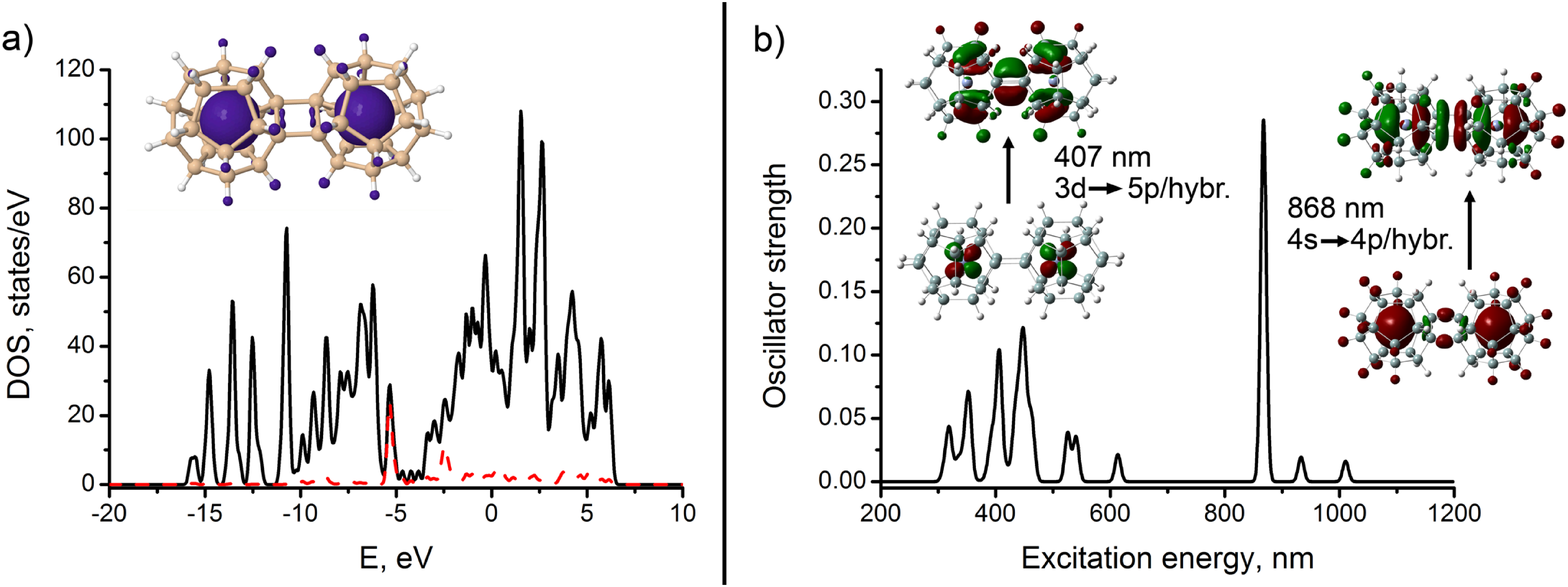}
\caption{a) Total DOS (black solid line) and DOS projected on metal dopants (red dashed line) for the (CrSi$_{20}$H$_{18}$)$_{2}$ dimer (highest occupied state lies at -4.67 eV, vacuum level is used as zero reference). The inset shows the spin density distribution within the structure; b) Calculated TD-DFT optical excitation spectrum of the (CrSi$_{20}$H$_{18}$)$_{2}$ dimer ground state, with the insets illustrating the dominant Kohn-Sham orbitals behind the characteristic Cr transitions.}
\label{fig4}
\end{figure*}

The analysis of the DOS diagram in \ref{fig4}a shows that the overall picture behind this high magnetization stays the same as discussed for the monomer CrSi$_{20}$H$_{20}$ case before. The distribution of the spin density, inset in \ref{fig4}a, further supports this view. Moreover, the optical excitation spectrum in \ref{fig4}b, computed within the time-dependent DFT (TD-DFT) linear response formalism \cite{gaussian03, tddft_details} indicates that the optical properties of the (CrSi$_{20}$H$_{18}$)$_{2}$ dimer still have a lot in common with those of the CrSi$_{20}$H$_{20}$ monomer, and even with an isolated Cr atom\cite{palagin12}. Similar to the monomer case, several well-defined excitations from Cr 4$s$ and 3$d$ orbitals can be clearly discerned, where due to the presence of two Cr atoms intensities are just typically higher than in the case of the monomer, and some of the peaks are split.

\begin{figure*}
\includegraphics[width=10cm]{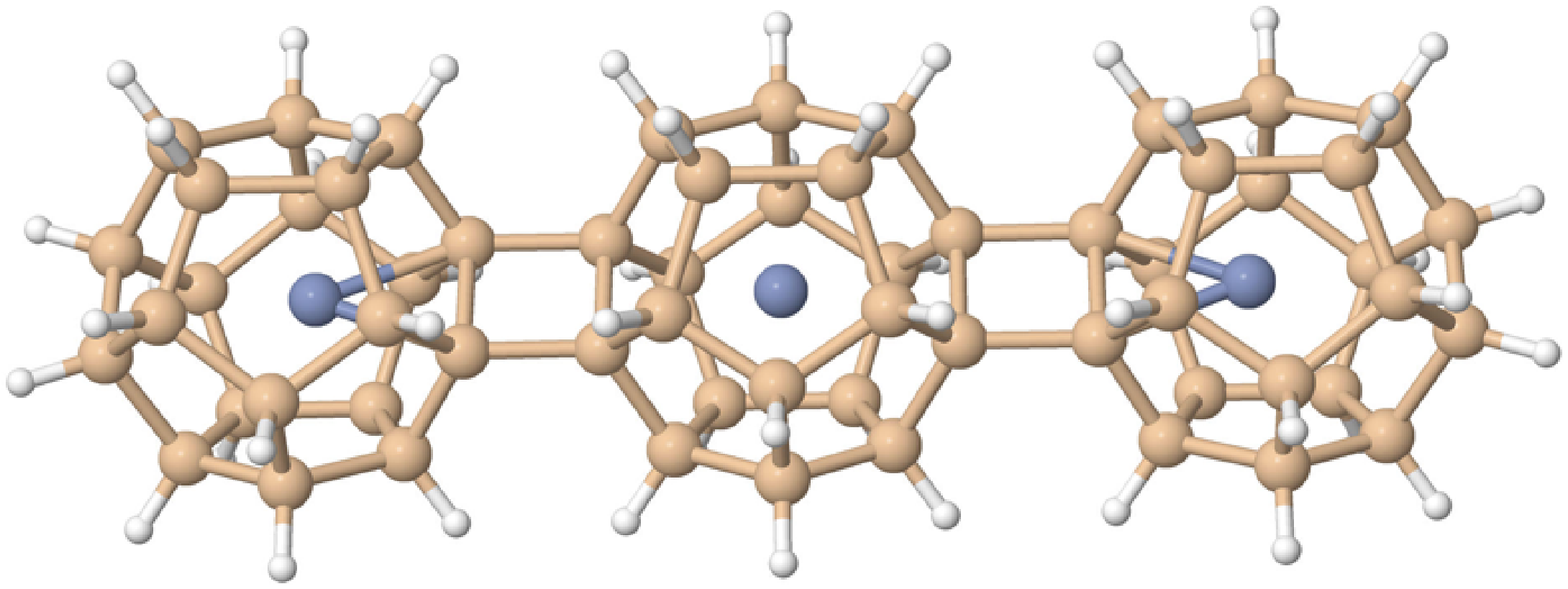}
\caption{Ground-state geometry of the [CrSi$_{20}$H$_{16}$](CrSi$_{20}$H$_{18}$)$_{2}$ trimer, featuring 18 unpaired electrons.}
\label{fig5}
\end{figure*}

Encouraged by these findings we have run a global geometry optimization of the linear \newline [CrSi$_{20}$H$_{16}$](CrSi$_{20}$H$_{18}$)$_{2}$ trimer, which is at the edge of present-day computational capabilities. The resulting ground-state geometry is presented in \ref{fig5} and indeed corresponds to a highly symmetrical, doubly Si-Si bridge bonded trimer structure with a very high spin moment of 18 unpaired electrons altogether, again located predominantly on the three Cr dopants. The energetic gap to the next low-lying isomer has increased to an incredible 7.16\,eV. We stress again that as a result of extensive sampling this includes the stability against decomposition into any set of separated smaller clusters containing in sum the same number of species. For instance, we have explicitly evaluated by separate global geometry optimization of the smaller fragments that the obtained [CrSi$_{20}$H$_{16}$](CrSi$_{20}$H$_{18}$)$_{2}$ trimer is by 8.15\,eV more stable than a set formed of two CrSi$_{20}$H$_{20}$, one CrSi$_{8}$ and one Si$_{12}$H$_{12}$, or by 9.03\,eV more stable than a set formed of two CrSi$_{20}$H$_{20}$, one Si$_{8}$ and one CrSi$_{12}$H$_{12}$.

\subsection{Routes to polymerization}

\begin{figure*}
\includegraphics[width=10cm]{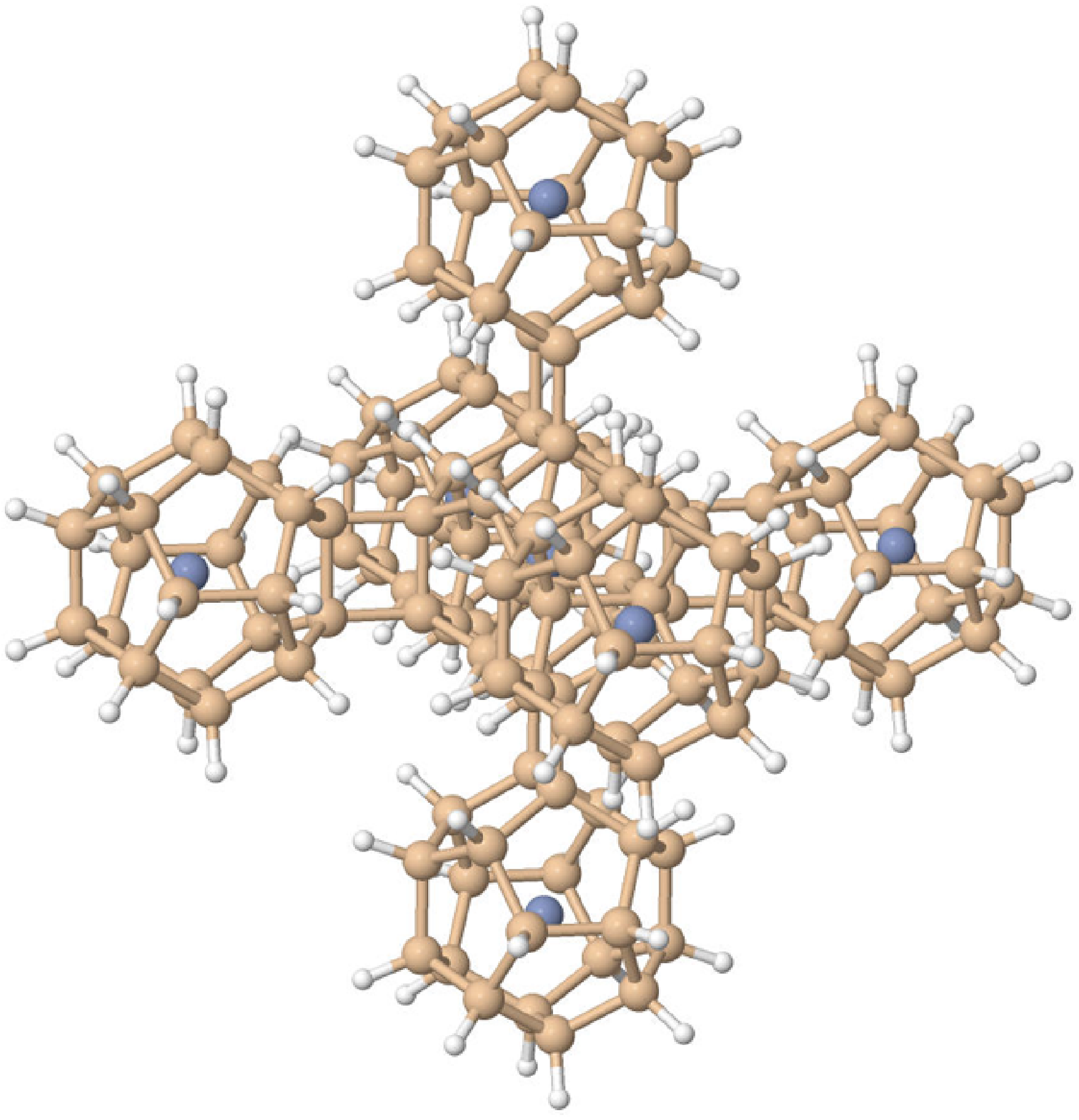}
\caption{Locally optimized geometry of a [CrSi$_{20}$H$_{8}$](CrSi$_{20}$H$_{18}$)$_{6}$ Si-doubly bridge bonded 3D octahedral aggregate, exhibiting 42 unpaired electrons.}
\label{fig6}
\end{figure*}

The dimerization and trimerization results indicate possibility of assembling very stable aggregates that largely conserve the magnetic properties of the dopant atoms. Involving cages with higher degree of dehydrogenation then opens up a pathway to more complicated network architectures, $e.g.$ the CrSi$_{20}$H$_{12}$ cage with four H vacancy pairs might be used for building two-dimensional sheets. Another important feature here is that the doubly-bridged Si bonding between the cages compensates for the H-vacancy pair induced Si undercoordination in the monomers. Whereas CrSi$_{20}$H$_{20-2n}$ monomers with larger numbers of H vacancy pairs were found to keep their structural integrity, the increased $M$-Si interaction still led to a quenching of the dopant spin moment, $cf.$ \ref{fig1}. In contrast, within the dimer and trimer structures the magnetic moment is reestablished and contributes to the high total number of unpaired electrons. Indeed, consistent with this understanding we have managed to obtain the locally optimized [CrSi$_{20}$H$_{8}$](CrSi$_{20}$H$_{18}$)$_{6}$ Si-doubly bridge bonded 3D octahedral aggregate shown in  \ref{fig6} with the overall number of unpaired electrons reaching an impressive 42. At the implied reminimized $M$-Si cage interaction within the aggregates our findings should also not be specific to the Cr dopant used as showcase. In our previous work we have established that dopants from the entire 3$d$ transition metal series yield $M$Si$_{20}$H$_{20}$ cages as ground state isomers.\cite{palagin12} Correspondingly, we expect the here suggested dehydrogenation route to polymerization to hold for a wider range of in particular magnetic dopants, which would then offer a flexible toolbox to engineer electronic or magnetic properties in hetero-aggregates.\cite{nakajima12}

In our view crucial for a potential synthesis is the structural completion of the individual endohedral cages in the doubly-bridge bonded aggregates. In this respect, they are real cluster-assembled materials and thus differ qualitatively from known Si clathrates which are also composed of endohedral fullerene-like building blocks ($e.g.$ $M$Si$_{20}$, $M$Si$_{24}$, $M$Si$_{28}$). Elements of a typical Si clathrate usually share facets \cite{mcmillan99, pouchard99, shevelkov04, faessler11}, such that all Si valence bonds are saturated by surrounding Si atoms. They can thus not be separated into structurally complete fullerene monomers and represent rather a space-filling framework than a cluster assembly at the molecular level. In contrast to concomitant clathrate synthesis routes $via$ controlled annealing of bulk material, we thus rather expect polymerization strategies $via$ controlled dehydrogenation of formed $M$Si$_{20}$H$_{20}$ clusters to potentially realize the here suggested aggregates. A central feature for this could be their very high stability, which renders $e.g.$ a simple H$_2$ abstraction reaction upon aggregation exothermic. For the dimer formation reaction CrSi$_{20}$H$_{20}$ + CrSi$_{20}$H$_{20}$ $\rightarrow$ (CrSi$_{20}$H$_{18}$)$_2$ + 2H$_2$ we compute this as 0.35\,eV (0.17\,eV per formed H$_2$) released upon aggregation, and similarly 0.14\,eV per formed H$_2$ for the abstraction of 4 resp. 12 H$_2$ molecules upon the formation of the trimer and 3D octahedral aggregate shown in \ref{fig5} and \ref{fig6}, respectively. This suggests a range of hydrogen chemical potential to selectively dehydrogenate and build corresponding oligomers, $e.g.$ in solution.

\subsection{Conclusions}

In summary, we have sytematically assessed the possibility of using endohedrally-doped hydrogenated Si-cages as building blocks for constructing highly magnetic materials. Our unbiased first-principles global geometry optimization highlights the structural integrity of the cage geometry of $e.g.$ the CrSi$_{20}$H$_{20}$ fullerene despite an increasing degree of dehydrogenation. The preferentially formed H-vacancy pairs then act as natural "docking sites" for polymerization. Depending on the amount of H-vacancy pairs of the involved monomers, this yields network architectures of any morphology, including linear chains, 2D sheets or 3D structures. 

In contrast to face-sharing Si clathrates, the resulting aggregates represent real cluster assembled materials, in which structurally complete endohedral Si cages are doubly bridged through Si-Si bonds. The latter geometric motif yields a high thermodynamic stability and has also been observed in preceding work on smaller $M$Si$_{16}$ clusters.\cite{torres11,nakajima12} For the present hydrogenated fullerenes, it affords a flexible and controllable molecular assembly, while maintaining the high magnetic moments of the dopant atoms inside the constitutent cluster monomers. Taking into account the possibility to also stabilize larger cages (such as Si$_{24}$H$_{24}$, Si$_{26}$H$_{26}$, Si$_{28}$H$_{28}$) and combining metal dopants having different spin states\cite{palagin12} in hetero-aggregates, this strongly suggests such assemblies of hydrogenated endohedral Si-clusters as promising candidates for the construction of highly magnetic nanostructured materials.

\section{Computational details}

All ground-state total energy calculations in this work have been performed with the all-electron full-potential DFT code FHI-aims\cite{blum09, ren12}. Electronic exchange and correlation was treated within the generalized-gradient approximation functional due to Perdew, Burke and Ernzerhof (PBE)\cite{perdew96}. For comparison the energetic differences between the found isomers were also systematically re-computed on the hybrid functional level with the PBE0 functional\cite{pbe0}, without ever obtaining any qualitative changes that would conflict with the conclusions deduced from the standard PBE calculations. All sampling calculations were done with the "tier2" atom-centered basis set using "tight" settings for numerical integrations. The stability of the identified minima has been confirmed by vibrational frequency analysis. For the subsequent electronic structure analysis of the optimized structures the electron density was recomputed with an enlarged "tier3" basis set\cite{blum09}. As detailed in our previous works\cite{gramzow10, palagin11, palagin12}, systematic convergence tests confirm full convergence of the settings with respect to the target quantities (energetic differences between isomers, total electron and spin density distributions). To ensure that the obtained geometries indeed represent the ground-state structures for all considered systems we relied on basin-hopping (BH) based global geometry optimization\cite{doye97, wales00}, which samples the potential energy surface (PES) through consecutive jumps from one local minimum to another. In our implementation\cite{gehrke09, gramzow10, palagin11, palagin12} it is achieved by random displacement of atoms in the cluster in a so-called trial move followed by a local geometry optimization. A Metroplis-type acceptance rule is used to either accept or reject the jump into the PES minimum reached by the trial move.

%% classes.  Rather than use \section, an appropriate macro is
%% provided that will always work.
%%%%%%%%%%%%%%%%%%%%%%%%%%%%%%%%%%%%%%%%%%%%%%%%%%%%%%%%%%%%%%%%%%%%%

\acknowledgement

Funding within the DFG Research Unit FOR1282 and support of the TUM Faculty Graduate Center Chemistry is gratefully acknowledged. We thank Saskia Stegmaier for fruitful discussions concerning clathrate structures.

%% \suppinfo macro.
%%%%%%%%%%%%%%%%%%%%%%%%%%%%%%%%%%%%%%%%%%%%%%%%%%%%%%%%%%%%%%%%%%%%%
\begin{suppinfo}

Detailed information about the structures, relative energies and energetics of frontier orbitals of the identified low-lying isomers can be found in the Supplemental Material.
\end{suppinfo}


\bibliography{palagin_cluster_aggregates}

\begin{mcitethebibliography}{32}
\bibitem[tdd()]{tddft_details}
To calculate the excitation spectrum of the cluster the first 125 transitions
  have been calculated using PBE0 functional and LANL2DZ\cite{lanl2dz} basis
  set within the TD-DFT\cite{tddft} approach.\relax
\mciteBstWouldAddEndPunctfalse
\mciteSetBstMidEndSepPunct{\mcitedefaultmidpunct}
{}{\mcitedefaultseppunct}\relax
\EndOfBibitem
\end{mcitethebibliography}

\end{document}